\documentclass[preprint,preprintnumbers,amssymb,prb,superscriptaddress]{revtex4}
\usepackage{graphicx}

\begin{document}

\title{Graphene: carbon in two dimensions}
\author{Mikhail I. Katsnelson}
\affiliation{Institute for Molecules and Materials, Radboud
University Nijmegen, 6525 ED Nijmegen, The Netherlands}

\begin{abstract}
{\bf Carbon is one of the most intriguing elements in the Periodic
Table. It forms many allotropes, some being known from ancient times
(diamond and graphite) and some discovered ten to twenty years ago
(fullerenes, nanotubes). Quite interestingly, the two-dimensional
form (graphene) has been obtained only very recently, and
immediately attracted great deal of attention. Electrons in
graphene, obeying linear dispersion relation, behave like massless
relativistic particles, which results in a number of very peculiar
electronic properties observed in this first two-dimensional
material: from an anomalous quantum Hall effect to the absence of
localization. It also provides a bridge between condensed matter
physics and quantum electrodynamics and opens new perspectives for
carbon-based electronics.}
\end{abstract}

\maketitle

\section{Two-dimensional form of carbon}

Carbon plays a unique role in nature. Actually, the formation of
carbon in stars as a result of merging of three $\alpha$-particles
is a crucial process providing existence of relatively heavy
chemical elements in the Universe~\cite{fowler}. The capability of
carbon atoms to form complicated networks~\cite{pauling} is a
fundamental fact of organic chemistry and the base for existence of
life, at least, in its known forms. Even elemental carbon
demonstrates unusually complicated behavior forming a number of very
different structures. Apart from diamond and graphite known from
ancient times recently discovered
fullerenes~\cite{fulleren1,fulleren2,fulleren3} and
nanotubes~\cite{nanotube} are in the focus of attention of
physicists and chemists now. Thus, only three-dimensional (diamond,
graphite), one-dimensional (nanotubes), and zero-dimensional
(fullerenes) allotropes of carbon have been known till recently. The
\textit{two-dimensional} form was conspicuously missing, resisting
any attempt of its experimental observation.

This elusive two-dimensional form of carbon has been named
graphene, and, ironically, is probably the best theoretically
studied carbon allotrope. Graphene - planar, hexagonal
arrangements of carbon atoms is the starting point in all
calculations on graphite, carbon nanotubes and fullerenes. At the
same time, numerous attempts to synthesize these two-dimensional
atomic crystals usually fail, ending up with nanometer-size
crystallites~\cite{Oshima}. Actually, these difficulties were not
surprising, in light of a common believe that truly
two-dimensional crystals (in contrast with numerous
\textit{quasi}-two-dimensional systems known before) cannot
exist~\cite{peierls1,peierls2,landau,LL,mermin}. Moreover, during
a synthesis, any graphene nucleation sites will have very large
perimeter-to-surface ratio, thus promoting collapse into other
carbon allotropes.

\subsection{Discovery of graphene}

\begin{figure}[t]
%h=here, t=top, b=bottom, p=separate figure page
\begin{center}\leavevmode
\includegraphics[width=1\linewidth]{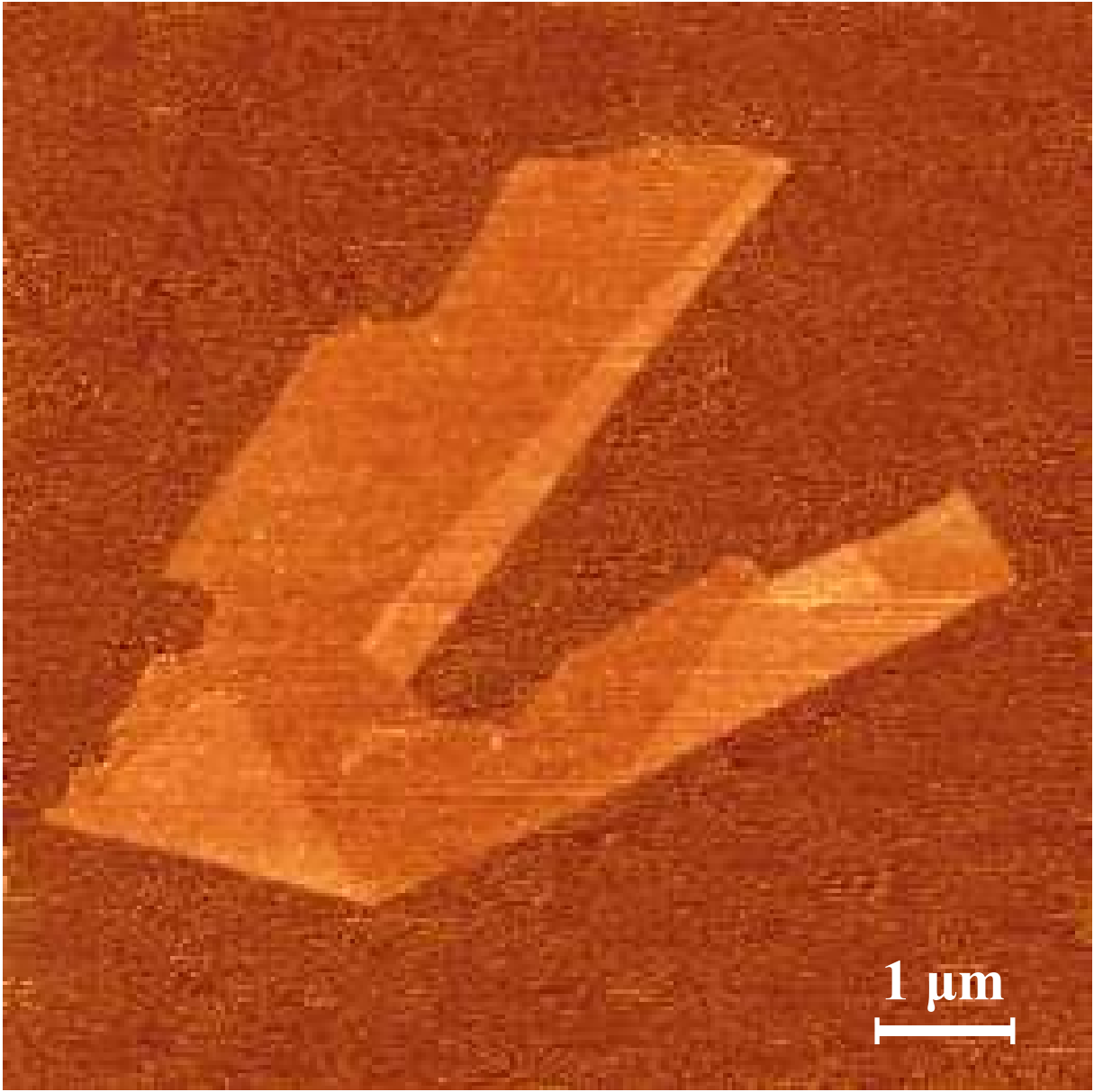}
\caption{Atomic force microscopy image of a graphene crystal on top
of oxidized silicon substrate. Folding of the flake can be seen. The
thickness of graphene measured corresponds to interlayer distance in
graphite.} \label{Flake}
\end{center}
\end{figure}

But in 2004 a group of physicists from Manchester University, led by
Andre Geim and Kostya Novoselov, used a very different and, at the
first glance, even naive approach to make a revolution in the field.
They started with three-dimensional graphite and extracted a single
sheet (a monolayer of atoms) from it by a technique called
micromechanical cleavage~\cite{kostya0,kostya1}, see
Fig.~\ref{Flake}. Graphite is a layered material and can be viewed
as consisting of a number of two-dimensional graphene crystals
weakly coupled together - exactly the property used by the
Manchester team. Moreover, by using this top-down approach and
starting with large three-dimensional crystals, the researchers from
Manchester avoided all the issues with the stability of small
crystallites. Furthermore, the same technique has been used by the
same group to obtain two-dimensional crystals of other
materials~\cite{kostya0}, such as boron-nitride, some
dichalcogenides and high-temperature superconductor Bi-Sr-Ca-Cu-O.
This astonishing finding sends an important message: two-dimensional
crystals do exist and they are stable at ambient conditions.

Amazingly, this humble approach allows easy production of high
quality, large (up to $100\;\mu m$ in size) graphene crystallites,
which immediately triggered enormous experimental
activity~\cite{kostya2,kim}. Moreover, the quality of the samples
produced are so good, that ballistic transport~\cite{kostya1} and
Quantum Hall Effect (QHE) can be easily observed~\cite{kostya2,kim}.
The former makes this new material a promising candidate for future
electronic applications (ballistic field effect transistor (FET)).
However, as the approach described definitely suits all the research
needs, other techniques, which would allow high yield of graphene,
are required for any industrial production. Among promising
candidates for this role one should mention exfoliation of
intercalated graphitic
compounds~\cite{Dresselhaus,Shioyama,Vicilis,Horiuchi,Stankovich}
and silicon sublimation from silicon-carbide substrates demonstrated
recently by Walt de Heer's group from Georgia Institute of
Technology~\cite{Berger}.

%\begin{figure}[t]
%h=here, t=top, b=bottom, p=separate figure page
%\begin{center}\leavevmode
%\includegraphics[width=1\linewidth]{CarbonAllotropes.eps}
%\caption{Crystal structures of different allotropes of carbon. Left
%to right: diamond and graphite (3D); graphene (2D); carbon nanotubes
%(1D); buckyballs (0D). Some crystal structures are adopted from
%"General Chemistry", Hill and Petrucci.} \label{CarbonAllotropes}
%\end{center}
%\end{figure}

\section{Stability issues in two-dimensions}

The fact that two-dimensional atomic crystals do exist, and
moreover, are stable under ambient conditions~\cite{kostya0} is
amazing by itself. According to the so called Mermin-Wagner
theorem~\cite{mermin}, there should be no long-range order in
two-dimensions, thus dislocations should appear in 2D crystals at
any finite temperature.

A standard description~\cite{born} of atomic motion in solids
assumes that amplitudes of atomic vibration $\bar{u}$ near their
equilibrium position are much smaller than interatomic distances
$d$; otherwise, the crystal would melt according to an empirical
Lindemann criterion (at the melting point, $\bar{u} \simeq 0.1
d$). Due to this smallness, thermodynamics of solids can be
successfully described in a picture of an ideal gas of phonons,
that is, quanta of atomic displacement waves (harmonic
approximation). In three-dimensional systems this view is
self-consistent in a sense that, indeed, fluctuations of atomic
positions calculated in the harmonic approximation turn out to be
small, at least, for low enough temperatures. In contrast, in a
two-dimensional crystal the number of long-wavelength phonons
diverge at low temperatures and, thus, amplitudes of interatomic
displacements calculated in the harmonic approximation are
diverging~\cite{peierls1,peierls2,landau}). According to similar
arguments, a flexible membrane embedded into the three-dimensional
space should be crumpled due to dangerous long-wavelength bending
fluctuations~\cite{nelson}.

However, it was demonstrated by efforts of theoreticians working
on soft-condensed matter during last twenty
years~\cite{nelson,peliti,radz} that these dangerous fluctuations
can be suppressed by anharmonic (nonlinear) coupling between
bending and stretching modes; as a result, the single-crystalline
membrane can exist but should be ``rippled''. This means arising
of ``roughness fluctuations'' with a typical height scaled with
the sample size $L$ as $L^{\zeta}$, with $\zeta \simeq 0.6$.
Indeed, the ripples are observed in graphene and play an important
role in their electronic properties~\cite{morozov}. However, these
investigations have just started (one can mention a few recent
works on Raman spectroscopy on graphene~\cite{Ferrari,Ensslin}),
and ``phononic'' aspects of two-dimensionality in graphene are
still very poorly understood.

Another important issue is a role of defects in thermodynamic
stability of two-dimensional crystals. Finite concentration of such
defects as dislocations and disclinations would destroy long-range
translational and orientational order, respectively. The detailed
analysis~\cite{nelson} shows that the dislocations in {\it flexible}
membranes have a finite energy (of order of cohesion energy
$E_{coh}$), due to screening of bending deformations, whereas the
energy of disclinations is logarithmically divergent with the size
of crystallite. This means that, rigorously speaking, the
translational long-range order (but not the orientational one) is
broken at any finite temperatures $T$. However, the density of
dislocations in the equilibrium is exponentially small for large
enough $E_{coh}$ (in comparison with the thermal energy $k_B T$) so,
in practice, this restriction is not very serious for strongly
bonded two-dimensional crystals like graphene.

         %\begin{figure}[t]
         %h=here, t=top, b=bottom, p=separate figure page
         %\begin{center}\leavevmode
         %\includegraphics[width=1\linewidth]{GraphiteBand6.eps}
         %\caption{Crystal and electronic structure of graphite.}
         %\label{GraphiteBand6}
         %\end{center}
         %\end{figure}

\begin{figure}[t]
%h=here, t=top, b=bottom, p=separate figure page
\begin{center}\leavevmode
\includegraphics[width=1\linewidth]{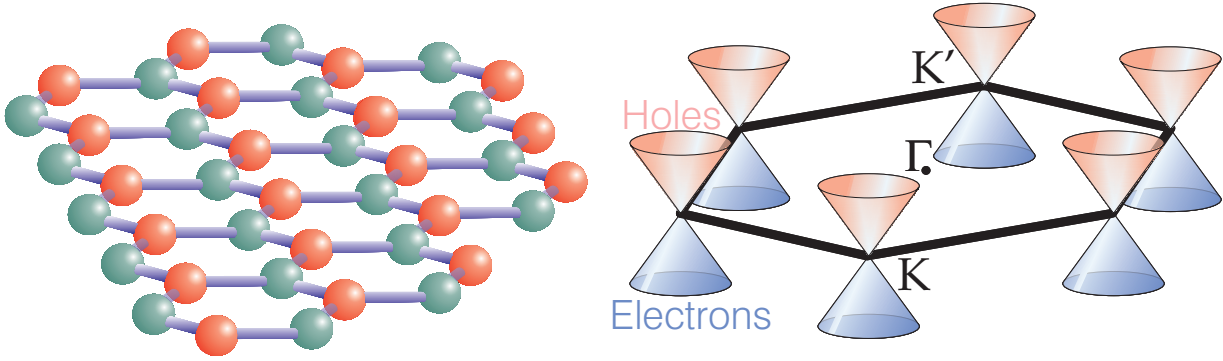}
\caption{Left: Crystallographic structure of graphene. Atoms from
different sublattices (A and B) are marked by different colors.
Right: Band structure of graphene in the vicinity of the Fermi
level. Conductance band touches valence band at $K$ and $K'$
points. } \label{GrapheneBand6}
\end{center}
\end{figure}

%\begin{figure}[t]
%h=here, t=top, b=bottom, p=separate figure page
%\begin{center}\leavevmode
%\includegraphics[width=1\linewidth]{GrapheneCrystalStructure.eps}
%\caption{Crystallographic structure of graphene. Atoms from
%different sublattices (A and B) are marked by different colours.}
%\label{GrapheneCrystalStructure}
%\end{center}
%\end{figure}

\section{Electronic structure of graphene: Linear dispersion relation}

The electronic structure of graphene follows already from simple
nearest-neighbor tight-binding approximation~\cite{wallace}.
Graphene has two atoms per unit cell, which results in two
``conical'' points per Brillouin zone with the band crossing, $K$
and $K'$. Near these crossing points the electron energy is linearly
dependent on the wave vector. Actually, this behavior follows from
symmetry considerations~\cite{slon} and, thus, is robust with
respect to long-range hopping processes
(Figure~\ref{GrapheneBand6}).

What makes graphene so attractive for research, is that the spectrum
closely resembles the Dirac spectrum for massless
fermions~\cite{semenoff,haldane}. The Dirac equation describes
relativistic quantum particles with spin 1/2, such as electrons. The
essential feature of the Dirac spectrum following from the basic
principles of quantum mechanics and relativity theory is the
existence of antiparticles. More specifically, states at positive
and negative energies (electrons and positrons) are intimately
linked (conjugated), being described by different components of the
same spinor wavefunction. This fundamental property of the Dirac
equation is often referred to as the charge-conjugation symmetry.
For Dirac particles with mass $m$ there is a gap between minimal
electron energy, $E_0 = mc^2$, and maximal positron energy, -$E_0$
($c$ is the speed of light). When the electron energy $E \gg E_0$
the energy is linearly dependent on the wavevector ${\bf k}$, $E = c
\hbar k$. For {\it massless} Dirac fermions, the gap is zero and
this linear dispersion law holds at any energies. For this case,
there is an intimate relation between the spin and motion of the
particle: spin can be only directed along the propagation direction
(say, for particles) or only opposite to it (for antiparticles). In
contrast, massive spin-1/2 particles can have {\it two} values of
spin projection on any axis. In a sense we have a unique situation
here: charged massless particles. Although it is a popular textbook
example - no such particles have been observed before.

The fact that charge carriers in graphene are described by the
Dirac-like spectrum, rather than the usual Schr\"{o}dinger
equation for nonrelativistic quantum particles, can be seen as a
consequence of graphene's crystal structure, which consists of two
equivalent carbon sublattices A and B (see the
Figure~\ref{GrapheneBand6}). Quantum mechanical hopping between
the sublattices leads to the formation of two energy bands, and
their intersection near the edges of the Brillouin zone yields the
conical energy spectrum. As a result, quasiparticles in graphene
exhibit a linear dispersion relation $E = \hbar k v_F$, as if they
were massless relativistic particles (for example, photons) but
the role of the speed of light is played here by the Fermi
velocity $v_F\approx c/300$. Due to the linear spectrum, one can
expect that graphene's quasiparticles behave differently from
those in conventional metals and semiconductors where the energy
spectrum can be approximated by a parabolic (free-electron-like)
dispersion relation.

\section{Chiral Dirac electrons}

Although the linear spectrum is important, it is not its only
essential feature. Above zero energy, the current carrying states
in graphene are, as usual, electron-like and negatively charged.
At negative energies, if the valence band is not full, its
unoccupied electronic states behave as positively charged
quasiparticles (holes), which are often viewed as a
condensed-matter equivalent of positrons. Note however that
electrons and holes in condensed matter physics are normally
described by separate Schr\"{o}dinger equations, which are not in
any way connected (as a consequence of the so called Seitz sum
rule~\cite{VK}, the equations should also involve different
effective masses). In contrast, electron and hole states in
graphene should be interconnected, exhibiting properties analogous
to the charge-conjugation symmetry in the quantum electrodynamics
(QED)~\cite{slon,semenoff,haldane}. For the case of graphene, the
latter symmetry is a consequence of its crystal symmetry because
graphene's quasiparticles have to be described by two-component
wavefunctions, which is needed to define relative contributions of
sublattices A and B in quasiparticles' make-up. The two-component
description for graphene is very similar to the one by spinor
wavefunctions in QED but the ``spin'' index for graphene indicates
sublattices rather than the real spin of electrons and is usually
referred to as pseudospin $\sigma$. This allows one to introduce
chirality~\cite{haldane}, that is formally a projection of
pseudospin on the direction of motion, which is positive and
negative for electrons and holes, respectively.

The description of electron spectrum of graphene in terms of Dirac
massless fermions is a kind of continuum-medium description
applicable for electron wavelengths much larger than interatomic
distance. However, even at these space scales there is still a
reminiscence of the structure of elementary cell, that is, existence
of two sublattices. In terms of the continuum field theory, this can
be described only as {\it internal} degree of freedom of charge
carriers which is just the chirality.

\begin{figure}[t]
%h=here, t=top, b=bottom, p=separate figure page
\begin{center}\leavevmode
\includegraphics[width=1\linewidth]{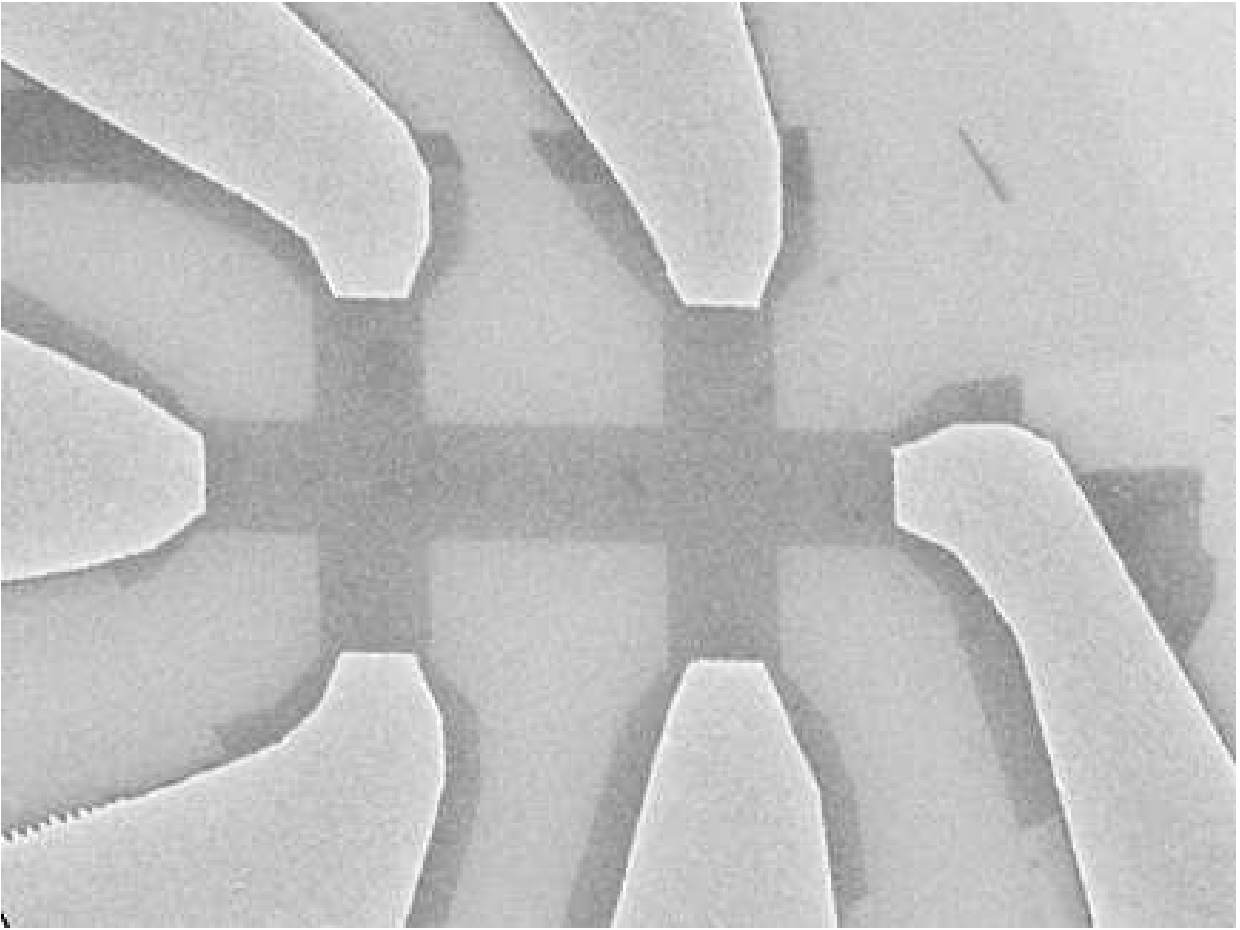}
\caption{SEM micrograph of a graphene device. Graphene crystal has
been connected by golden electrodes and patterned into Hall bar
geometry by e-beam lithography with subsequent reactive plasma
etching. The width of the channel is $1\mu m$. (Courtesy of K.
Novoselov and A. Geim)}. \label{Sample}
\end{center}
\end{figure}

This description is based on oversimplified nearest-neighbor
tight-binding model. However, it was proven experimentally that
charge carriers in graphene do have this Dirac-like gapless energy
spectrum~\cite{kostya2,kim}. This was demonstrated in transport
experiments (Fig.~\ref{Sample}) via investigation of Schubnikov - de
Haas effect, that is, oscillations of resistivity in high magnetic
fields at low temperatures.

\section{Anomalous Quantum Hall Effect in graphene}

\begin{figure}[t]
%h=here, t=top, b=bottom, p=separate figure page
\begin{center}\leavevmode
\includegraphics[width=1\linewidth]{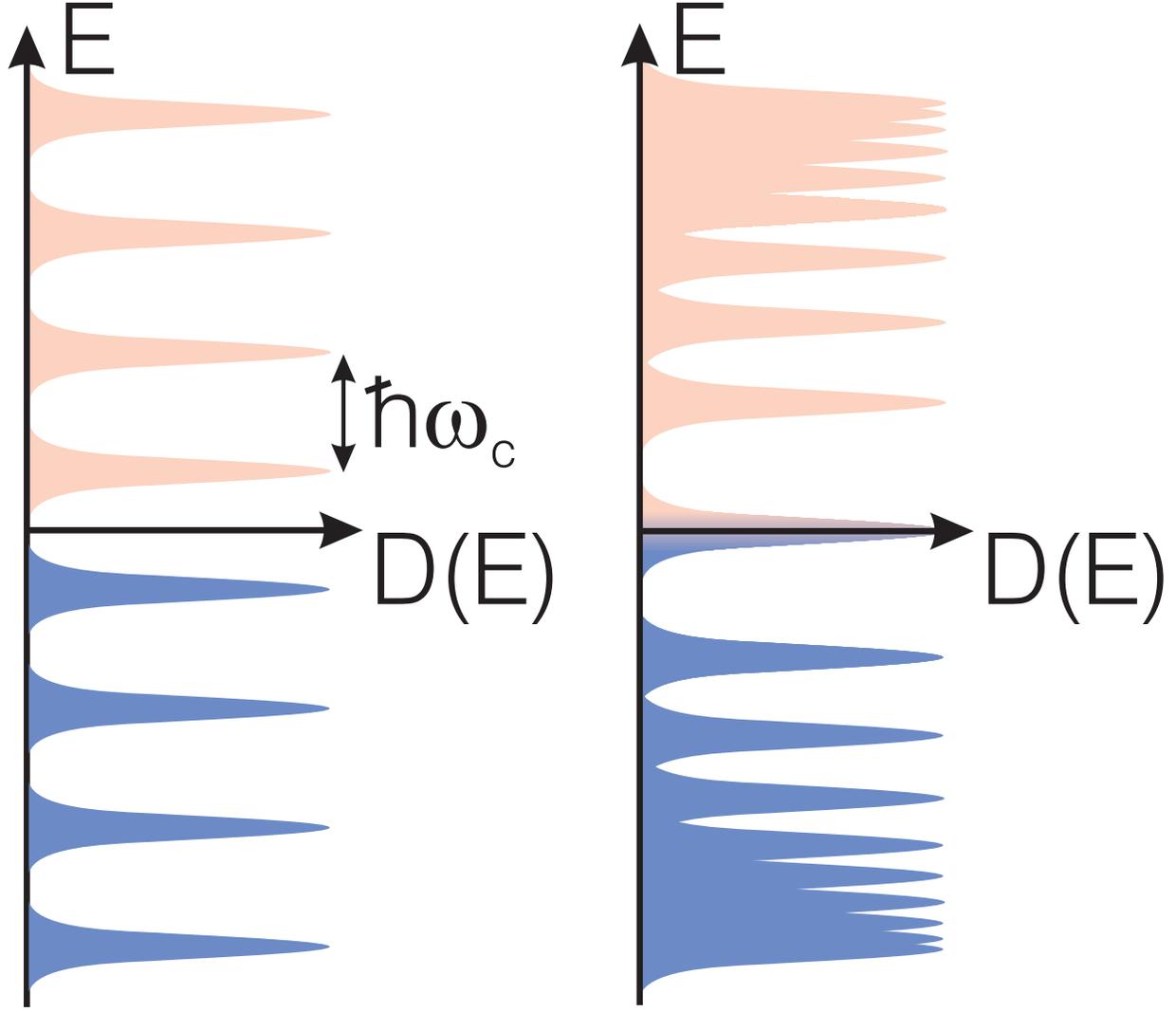}
\caption{Landau levels for Shr\"{o}dinger electrons with two
parabolic bands touching each other at zero energy (left panel).
Landau levels for Dirac electrons (right panel).}
\label{LandauLevels}
\end{center}
\end{figure}

Magnetooscillation effects such as de Haas - van Alphen
(oscillations of the magnetization) or Schubnikov - de Haas effect
(magneto-oscillations in the resistance) are among the most
straightforward and reliable tools to investigate electron energy
spectrum in metals and semiconductors~\cite{Ashcroft}. In
two-dimensional systems with a constant magnetic field ${\bf B}$
perpendicular to the system plane the energy spectrum is discrete
(Landau quantization). For the case of massless Dirac fermions the
energy spectrum takes the form (see, e.g.,~\cite{GS})
\begin{equation}
E_{\nu \sigma }= \pm \sqrt{2\left| e\right| B\hbar v_{F}^{2}\left(
\nu +1/2\pm 1/2\right) }
\end{equation}
where $v_{F}$ is the electron velocity, $\nu = 0,1,2...$ is the
quantum number and the term with $\pm 1/2$ is connected with the
chirality (Figure~\ref{LandauLevels}). Just to remind that in the
usual case of parabolic dispersion relation the Landau level
sequence is $E=\hbar\omega_c(\nu +1/2)$ where $\omega_c$ is the
frequency of electron rotation in the magnetic field (cyclotron
frequency)~\cite{Ashcroft}.

\begin{figure}[t]
%h=here, t=top, b=bottom, p=separate figure page
\begin{center}\leavevmode
\includegraphics[width=1\linewidth]{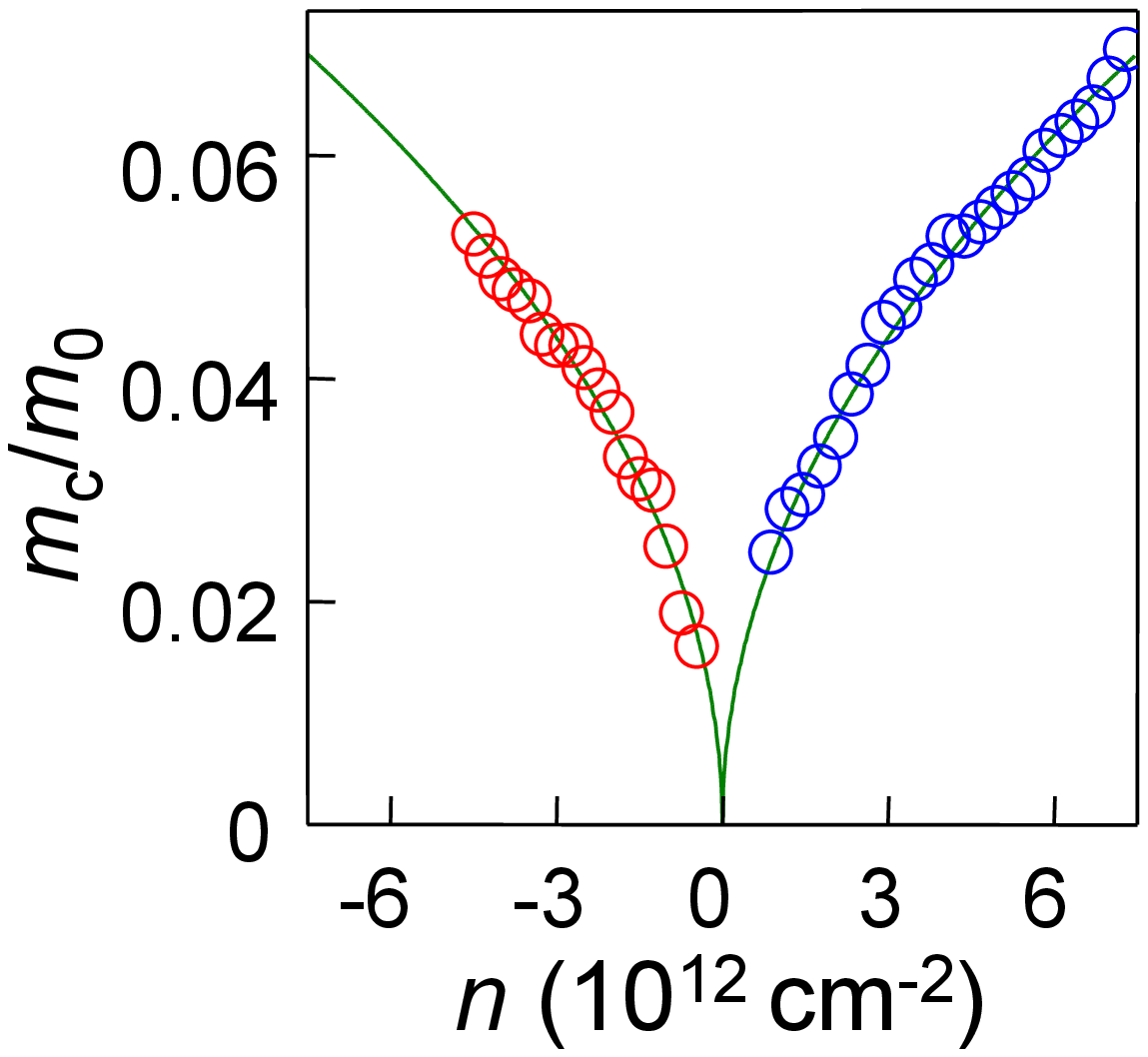}
\caption{Electrons and holes cyclotron mass as a function of carrier
concentration in graphene. The square-root dependence suggests
linear dispersion relation.} \label{CyclotronMass}
\end{center}
\end{figure}

By changing the value of magnetic field at a given electron
concentration (or, vice versa, electron concentration for a given
magnetic field) one one can tune the Fermi energy $E_{F}$ with one
of the Landau levels. Such crossing changes drastically all
properties of metals (or semiconductors) and, thus, different
physical quantities will oscillate with the value of the inverse
magnetic field. By measuring the period of these oscillations
$\Delta \left( 1/B\right) $ we obtain an information about the area
$\mathcal{A}$ inside the Fermi surface (for two-dimensional systems,
this area is just proportional to the charge-carrier concentration
$n$). The amplitude of the oscillations allows us to measure the
effective cyclotron mass which is proportional to $\partial
\mathcal{A}/\partial E_{F}$~\cite{VK,Ashcroft}. For the case of
massless Dirac fermions (linear dependence of the electron energy on
its momentum) this quantity should be proportional to $\sqrt{n}$
which was exactly the behavior observed simultaneously by the
Manchester group and a group of Philip Kim and Horst Stormer from
Columbia University~\cite{kostya2,kim} (see
Figure~\ref{CyclotronMass}).

\begin{figure}[t]
%h=here, t=top, b=bottom, p=separate figure page
\begin{center}\leavevmode
\includegraphics[width=0.5\linewidth]{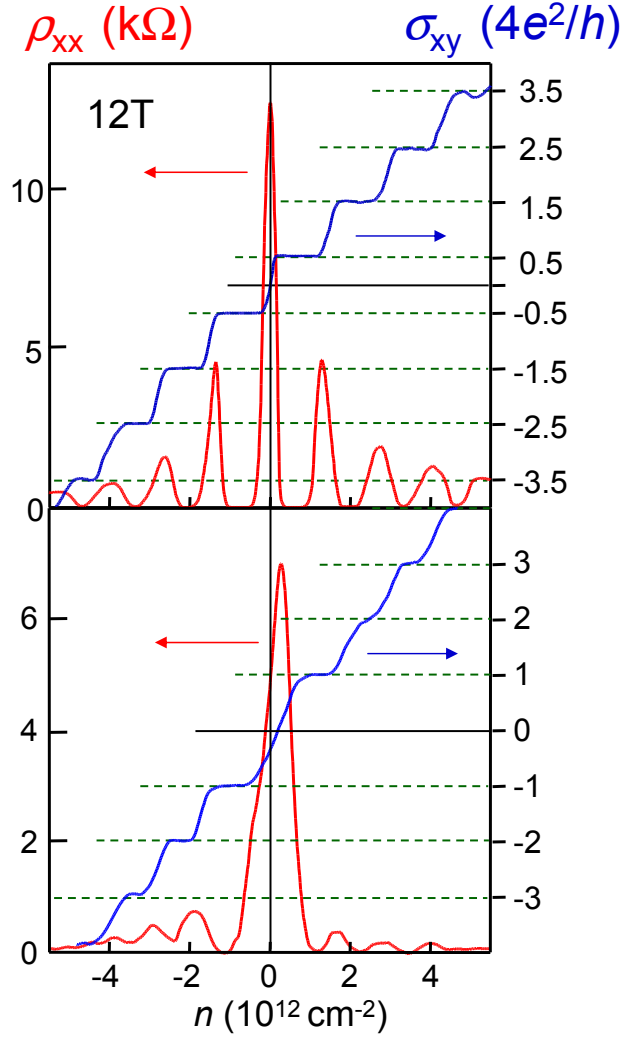}
\caption{Resistivity (red curves) and Hall conductivity (blue
curves) as a function of carrier concentration in graphene (upper
panel) and bi-layer graphene (lower panel).} \label{QHE}
\end{center}
\end{figure}

An important peculiarity of the Landau levels for massless Dirac
fermions is the existence of zero-energy states (with $\nu =0$ and
minus sign in equation (1)). This situation differs essentially from
usual semiconductors with parabolic bands where the first Landau
level is shifted by $1/2\;\hbar\omega_c$.  As it has been shown by
the Manchester and the Columbia groups~\cite{kostya2,kim}, the
existence of the zero-energy Landau level leads to an anomalous
quantum Hall effect (QHE) with \textit{half-integer} quantization of
the Hall conductivity (Fig.~\ref{QHE}, upper panel), instead of
\textit{integer} one (for a review of the QHE, see, e.g.,
Ref.~\cite{QHE}). Usually, all Landau levels have the same
degeneracy (a number of electron states with a given energy) which
is just proportional to the magnetic flux through the system. As a
result, the plateaus in the Hall conductivity corresponding to the
filling of first $\nu$ levels are just integer (in units of the
conductance quant $e^{2}/h$). For the case of massless Dirac
electrons, the zero-energy Landau level has twice smaller degeneracy
than any other level (it corresponds to the minus sign in the
equation (1) whereas $p$-th level with $p\geq 1$ are obtained twice,
with $\nu =p$ and minus sign, and with $\nu =p-1$ and plus sign). A
discovery of this anomalous QHE was the most direct evidence of the
Dirac fermions in graphene~\cite{kostya2,kim}.

\subsection{Index theorem}

The deepest view on the origin of zero-energy Landau level and thus
anomalous QHE is provided by a famous Atiyah-Singer index theorem
which plays an important role in modern quantum field theory and
theory of superstrings~\cite{index}. The Dirac equation has a
charge-conjugation symmetry between electrons and holes. This means
that for any electron state with a positive energy $E$ a
corresponding conjugated hole state with the energy $-E$ should
exist. However, the states with zero energy can be, in general,
anomalous. For curved space (e.g., for a deformed graphene sheet
with some defects of crystal structure) and/or in the presence of so
called ``gauge fields'' (usual electromagnetic field provides just a
simplest example of these fields) sometimes an existence of states
with zero energy is guaranteed by topological reasons, this states
being chiral (in our case this means that depending on the sign of
the magnetic field there is only sublattice A or sublattice B states
which contribute to the zero-energy Landau level). This means, in
particular, that the number of these states expressed in terms of
total magnetic flux is a topological invariant and remains the same
even if the magnetic field is inhomogeneous~\cite{kostya2}. This is
an important conclusion since the ripples on graphene create
effective inhomogeneous magnetic fields of magnitude up to 1T
leading, in particular, to suppression of the weak
localization~\cite{morozov}. However, due to these topological
arguments they cannot destroy the anomalous QHE in graphene. For
further insight in to the applications of the index theorem to
two-dimensional systems, and, in particular, to graphene see
Refs.~\cite{annphys,stone}.

\subsection{Quasi-classical consideration: Berry phase.}

An alternative view on the origin of the anomalous QHE in graphene
is based on the concept of ``Berry phase''~\cite{berry}. Since the
electron wave function is a two-component spinor, it has to change
sign when the electron moves along the close contour. Thus the
wave function gains an additional phase $\phi = \pi$. In
quasiclassical terms (see, e.g., Refs.~\cite{VK,mikitik}),
stationary states are nothing but electron standing waves and they
can exist if the electron orbit contains, at least, half of the
wavelength. Due to additional phase shift by the Berry phase, this
condition is satisfied already for the zeroth length of the orbit,
that is, for zero energy!

Other aspects of the QHE in graphene are considered in
papers~\cite{levitov,gus,per,cas}.

\section{Anomalous Quantum Hall Effect in bilayer graphene}

In relativistic quantum mechanics, chirality is intimately connected
with relativistic considerations which dictates, at the same time,
linear energy spectrum for massless particles. Discovery of graphene
opens a completely new opportunity, to investigate \textit{chiral}
particles with \textit{parabolic (non-relativistic)} energy
spectrum! This is the case of {\it bilayer} graphene~\cite{bilayer}.
For two carbon layers, the nearest-neighbor tight-binding
approximation predicts a gapless state with {\it parabolic} touching
in $K$ and $K'$ points, instead of conical one~\cite{falko,bilayer}.
More accurate consideration~\cite{peeters} gives a very small band
overlap (about 1.6 meV) but at larger energies the bilayer graphene
can be treated as the gapless semiconductor. At the same time, the
electron states are still characterized by chirality and by the
Berry phase (which is equal, in this case, $2\pi$ instead of $\pi$).
Exact solution of the quantum mechanical equation for this kind of
spectrum in a presence of homogeneous magnetic field gives the
result~\cite{falko,bilayer} $E_{\nu} \propto \sqrt{\nu \left( \nu
-1\right)}$ and, thus, the number of states with zero energy is
twice larger than for the case of monolayer ($\nu =0$ and $\nu =1$).
As a result, the QHE for bilayer graphene differs from both
single-layer one and conventional semiconductors, as it was found
experimentally~\cite{bilayer} by the Manchester team
(Fig.~\ref{QHE}, lower panel).

\section{Tunneling of chiral particles}

Chiral nature of electron states in the bilayer (as well as for the
case of single-layer graphene) is of crucial importance for electron
tunnelling through potential barriers and, thus, physics of
electronic devices such as ``carbon transistors''~\cite{ktsn}.

\subsection{Quantum tunneling}

\begin{figure}[t]
%h=here, t=top, b=bottom, p=separate figure page
\begin{center}\leavevmode
\includegraphics[width=0.5\linewidth]{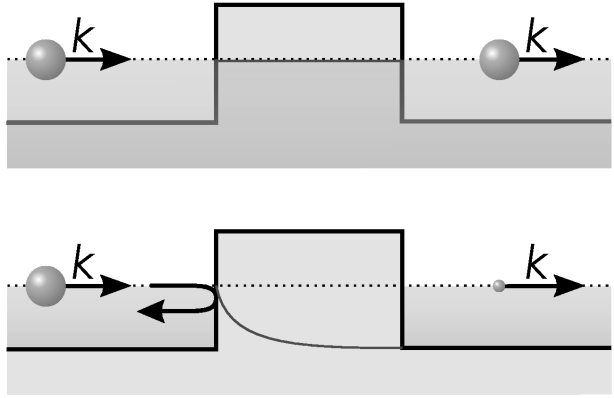}
\caption{Tunneling in graphene (top panel) and in conventional
semiconductor (lower panel). The amplitude of electron wavefunction
(red line) remains constant in case of graphene and drops
exponentially for conventional tunneling. The size of the ball
indicate the amplitude of the incident and transmitted
wavefunction.} \label{KleinBarrier}
\end{center}
\end{figure}

Quantum tunneling is a consequence of very general laws of quantum
mechanics, such as the famous Heisenberg uncertainty relations. A
classical particle cannot propagate through a region where its
potential energy is higher than its total energy
(figure~\ref{KleinBarrier}). However, due to the uncertainty
principle, for the case of quantum particle it is impossible to know
exact values of its coordinate and velocity, and, thus, its kinetic
and potential energy at the same time instant. Therefore,
penetration through the ``classically forbidden'' region turns out
to be possible. This phenomenon is widely used in modern electronics
starting from pioneering works by L. Esaki~\cite{esaki}.

\subsection{Klein paradox in graphene}

When potential barrier is smaller than a gap separating electron and
hole bands in semiconductors, the penetration probability decays
exponentially with the barrier height and width. Otherwise, a
resonant tunneling possible when the energy of propagating electron
coincides with one of the hole energy levels inside the barrier.
Surprisingly, in the case of graphene the transmission probability
for normally incident electrons is always equal to unity,
irrespective to the height and width of the
barrier~\cite{ktsn,Ferreira,FC}. In terms of quantum
electrodynamics, this behavior is related to the phenomenon known as
the Klein paradox~\cite{klein,dombey,pencil,ktsn}. This term usually
refers to a counter intuitive relativistic process in which an
incoming electron starts penetrating through a potential barrier if
its height exceeds twice the electron's rest energy $mc^2$. In this
case, the transmission probability $T$ depends only weakly on the
barrier height, approaching the perfect transparency for very high
barriers, in stark contrast to the conventional, nonrelativistic
tunnelling. This relativistic effect can be attributed to the fact
that a sufficiently strong potential, being repulsive for electrons,
is attractive for positrons and results in positron states inside
the barrier, which align in energy with the electron continuum
outside. Matching between electron and positron wavefunctions across
the barrier leads to the high-probability tunnelling described by
the Klein paradox. In other words, it reflects an essential
difference between nonrelativistic and relativistic quantum
mechanics. In the former case, we can measure accurately either
position of the electron or its velocity but not both of them
simultaneously. In the relativistic quantum mechanics, we cannot
measure even electron position with arbitrary accuracy since when we
try to do this we create electron-positron pairs from the vacuum and
we cannot distinguish our original electron from these newly created
electrons. Graphene opens a way to investigate this counterintuitive
behavior in a relatively simple bench-top experiment, whereas
originally it was connected with only some very exotic phenomena
such as collisions of ultraheavy nuclei or black hole evaporations
(for more references and explanations, see~\cite{ktsn,pencil}).

\subsection{Tunneling of chiral quasiparticles in bilayer graphene}

\begin{figure}[t]
%h=here, t=top, b=bottom, p=separate figure page
\begin{center}\leavevmode
\includegraphics[width=0.5\linewidth]{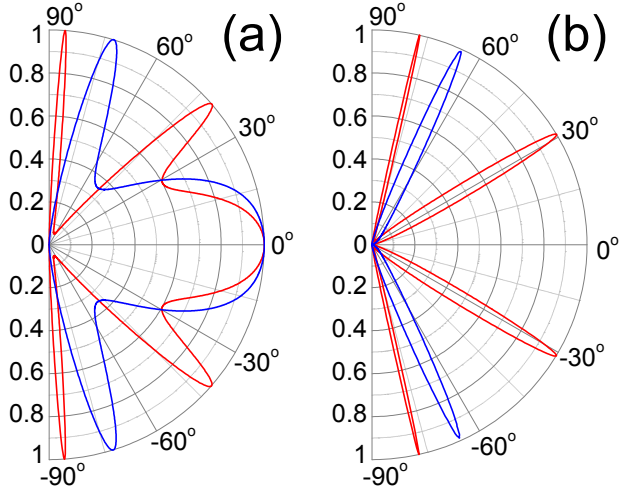}
\caption{Transmission probability $T$ through a $100$-nm-wide
barrier as a function of the incident angle for single- (a) and
bi-layer (b) graphene. The electron concentration n outside the
barrier is chosen as $0.5\times 10^{12}\;$cm$^{-2}$ for all cases.
Inside the barrier, hole concentrations p are $1\times 10^{12}$ and
$3\times 10^{12}\;$cm$^{-2}$ for red and blue curves, respectively
(such concentrations are most typical in experiments with graphene).
This corresponds to the Fermi energy $E$ of incident electrons
$\approx 80$ and $17\;$meV for single- and bi-layer graphene,
respectively, and $\lambda\approx50\;$nm. The barrier heights are
(a) $200$ and (b) $50\;$meV (red curves) and (a) $285$ and (b)
$100\;$meV (blue curves).} \label{KleinAngular}
\end{center}
\end{figure}

>From the point of view of possible applications it is a rather bad
news since it means that the ``carbon transistor'' from the
single-layer graphene cannot be closed by any external gate
voltage. On the contrary, it was shown in Ref.~\cite{ktsn} that
the chiral tunnelling for the case of {\it bilayer} leads to even
stronger suppression of the normally incident electron penetration
(Figure~\ref{KleinAngular}) than in conventional semiconductors.
It means that by creating a potential barrier (with external gate)
one can manipulate the transmission probability for ballistic
electrons in bilayer graphene. At the same time, there is always
some ``magic angle'' where the penetration probability equals
unity (Figure~\ref{KleinAngular}), which should be also taken into
account for a design of future carbon-based electronic devises.

\subsection{Absence of localization}

The discussed tunneling anomalies in single- and bilayer graphene
systems are expected to play an important role in their transport
properties, especially in the regime of low carrier concentrations
where disorder induces significant potential barriers and the
systems are likely to split into a random distribution of p-n
junctions. In conventional 2D systems, strong enough disorder
results in electronic states that are separated by barriers with
exponentially small transparency~\cite{ziman,lifshitz}. This is
known to lead to the Anderson localization. In contrast, in both
graphene materials all potential barriers are rather transparent at
least for some angles which does not allow charge carriers to be
confined by potential barriers that are smooth on atomic scale.
Therefore, different electron and hole ``puddles'' induced by
disorder are not isolated but effectively percolate, thereby
suppressing localization. This consideration can be important for
the understanding of the minimal conductivity $\approx e^2/h$
observed experimentally in both single-layer~\cite{kostya2} and
bilayer~\cite{bilayer} graphene. Further discussion of this minimal
conductivity phenomenon in terms of quantum relativistic effects can
be found in Refs.~\cite{zitter1,zitter2,carlo}.

%\begin{figure}[t]
%h=here, t=top, b=bottom, p=separate figure page
%\begin{center}\leavevmode
%\includegraphics[width=1\linewidth]{Superconductor.eps}
%\caption{An SEM micrograph of graphene crystal connected by
%superconducting electrodes. Supercurrent  due to proximity effect
%has been observed recently by Delft
%group~\cite{HubertSuperconductivity}. The gap between the electrodes
%is $70\;nm$} \label{Superconductor}
%\end{center}
%\end{figure}

\section{Graphene devices}

The unusual electronic properties of this new material make it a
promising candidate for future electronic applications. The
typical mobility easily achieved at the current state of
``graphene technology'' is around $20.000\;cm^2/V\cdot  s$, which
is already an order of magnitude higher than that of modern
silicon transistors, and it continue to grow as the quality of
graphene samples is improved. This insures ballistic transport on
sub-micron distances - the holy grail for any electronic engineer.
Probably the best candidates for graphene-based field effect
transistors will be devices based on quantum dots and those
utilizing $p-n$ junctions in bilayer graphene~\cite{ktsn,Nilsson}.

Another promising direction of investigation is spin-valve
devices. Due to negligible spin-orbit coupling, spin polarization
in graphene survives over submicron distances, which recently has
allowed observation of spin-injection and spin-valve effect in
this material~\cite{Hill}. It also has been shown by the group of
A. Morpurgo from Delft University that superconductivity can be
induced in graphene due to proximity
effect~\cite{HubertSuperconductivity}. Moreover, the value of
supercurrent can be controlled by an external gate voltage, which
can lead to the creation of superconducting FET.

Whenever the applications mentioned are a matter of further
investigations, there are some areas where graphene can be used
straightaway. Gas sensors are one of them. It has been shown by the
Manchester group that graphene can absorb gas molecules from the
surrounding atmosphere, which results in doping of the graphene
layer with electrons or holes, depending on the nature of the
absorbed gas~\cite{Schedin}. By monitoring changes of resistivity
one can sense minute concentrations of certain gases present in the
environment. Moreover, the sensitivity is so high that one can
detect an individual event of a single molecule attaching to the
surface of graphene gas sensor~\cite{Schedin}.

\section{Conclusions}

It is impossible, in this short paper, to review all aspects of
graphene physics and chemistry. We hope, however, that the
examples considered above already demonstrate its great interest
for both fundamental research (a new, unexpected bridge between
condensed matter and quantum field theory) and possible
applications. First of all, graphene is the first example of truly
two-dimensional crystals, in contrast with numerous {\it
quasi}-two-dimensional crystals known before. This opens many
interesting questions concerning thermodynamics, lattice dynamics
and structural properties of such systems. Further, being a
gapless semiconductor with linear energy spectrum the single-layer
graphene provides a realization of two-dimensional massless Dirac
fermion system which is of crucial importance for understanding
unusual electronic properties, such as anomalous QHE, absence of
the Anderson localization, etc. The bilayer graphene has a very
unusual gapless parabolic spectrum giving an example of the system
with electron wave equation different from both Dirac and
Schr\"{o}dinger ones. These peculiarities are important for
development of new electronic devises such as carbon transistors.

\textit{Acknowledgements}. I am thankful to Kostya Novoselov and
Andre Geim for many helpful discussions. This work was supported
by the Stichting voor Fundamenteel Onderzoek der Materie (FOM),
the Netherlands.

\end{document}